\documentclass[prb,twocolumn,showpacs]{revtex4}
\usepackage{graphicx}
\usepackage{amsmath}
\usepackage{amsxtra}
\usepackage{amssymb}
\usepackage{dcolumn}
\usepackage{float}
\usepackage{bm}
\usepackage[breaklinks=true,colorlinks,citecolor=blue,linkcolor=blue,urlcolor=blue]{hyperref}

\renewcommand{\Re}{\mathrm{Re}\,}
\renewcommand{\Im}{\mathrm{Im}\,}

\renewcommand{\Re}{\mathrm{Re}\,}
\renewcommand{\Im}{\mathrm{Im}\,}

\DeclareMathAlphabet{\bi}{OML}{cmm}{b}{it}
\def\be{\begin{equation}}
\def\ee{\end{equation}}
\def\bearr{\begin{eqnarray}}
\def\eearr{\end{eqnarray}}
\def\la{\langle}
\def\ra{\rangle}

\def\bs{\boldsymbol}

\begin{document}
\title{ Wave packet dynamics 
in various two-dimensional systems: a unified description}

\author{Ashutosh Singh}
\affiliation{Department of Physics, Indian Institute of Technology-Kanpur, Kanpur 208016, India}
\author{Tutul Biswas}
\affiliation{Department of Physics, Indian Institute of Technology-Kanpur, Kanpur 208016, India}
\author{Tarun Kanti Ghosh}
\affiliation{Department of Physics, Indian Institute of Technology-Kanpur, Kanpur 208016, India}
\author{Amit Agarwal}
\email{amitag@iitk.ac.in}
\affiliation{Department of Physics, Indian Institute of Technology-Kanpur, Kanpur 208016, India}

\date{\today}

\begin{abstract} 
In this article we present an exact and unified description of wave-packet dynamics in various
2D systems in presence of a transverse magnetic field. 
We consider an initial  minimum-uncertainty Gaussian wave-packet,
and  find that its long term dynamics  displays the universal phenomena
of spontaneous collapse and quantum revival. We estimate the timescales
associated with these phenomena based on very general arguments for various
materials, whose carrier dynamics is described either by the Schr\"odinger equation
or by the Dirac equation.

\end{abstract}



\maketitle
 
\section{INTRODUCTION} 

Spontaneous collapse and consequent quantum revival~[\onlinecite{parker, Perelman}]  occurs in the long
term dynamics of an injected wave-packet  in systems with non-equidistant energy levels due to quantum
interference. It has been investigated in a wide class of systems [\onlinecite{Robinett}] and the phenomenon
of wave packet collapse, revivals, and fractional revivals have been observed  experimentally in a number
of atomic and molecular systems [\onlinecite{stroud,stroud1,ewart,wals}]. However this phenomena of
purely quantum mechanical in origin,  is relatively less explored in condensed matter systems with
discrete Landau  energy levels, despite the fact that unlike {\it zitterbewegung} in solid state
systems [\onlinecite{zbS1,spinzb1,Zawadzki-review}] these oscillations are large and slow enough
for an experimental probe. 

{\it Zitterbewegung} and wave-packet dynamics in several 2D condensed matter systems with
Landau-levels have been explored
earlier [\onlinecite{green, zbgrph2,  Zawadzki_PRD_2010, Singh2014, Zawadzki-review, Schliemann, tutul}],
but the phenomena of spontaneous collapse and revival has largely gone
unaddressed [\onlinecite{Romera_PRB, Kramer, Romera2}].   Motivated by the
unified description of {\it zitterbewegung} in solid state systems
in Refs.~[\onlinecite{zbgen1,zbgen2}], in this article we present an exact
and unified description of the quantum wave-packet dynamics 
and the phenomena of collapse and quantum revival in various two
dimensional (2D) solid state systems. Initially when a well localized
wave packet is injected into a 2D system with 
non-equidistant Landau energy levels, it undergoes cyclotron
motion and evolves quasi-classically 
with periodicity $\tau_{\rm cl}$, for a number of cycles, with its
probability density spreading around the quasi-classical
trajectory. Non-equidistant nature of the discrete energy spectrum
then leads to destructive quantum interference and consequently the 
collapse of the wave-packet. The (almost) collapsed wave-packet regain
their initial waveform  and oscillate again with the quasi-classical
periodicity on a much longer time scale known as revival time ($\tau_{\rm rev} \gg
\tau_{\rm cl}$).
In addition, there is also the possibility of fractional revivals which
occurs at rational fraction of the revival time $\tau_{\rm rev}$ 
when the initial wave-packet evolves into a collection of mini wave-packets
resembling the waveform of the injected wave-packet [\onlinecite{Robinett, Perelman}].

In this article we consider several 2D systems, which can be classified  into two categories
based on the dynamics of the charge carriers: Schr\"odinger-like (non-relativistic
dynamics described by the Schr\"odinger equation) and Dirac-like
fermionic systems [\onlinecite{DM}] (relativistic dynamics described by the
Dirac equation). The so called `Schr\"odinger-like' materials include 
2D electron/hole gas (2DEG/2DHG) trapped at the interface of III-V semiconductor
hetero-structures, like AlGaAs-GaAs, and these
typically have a linear [\onlinecite{Gossard, Nitta}] or cubic spin orbit interaction
(SOI) terms [\onlinecite{hole1,Winkler,hole3,hole4, Sr1, Sr3}] in addition to the parabolic
dispersion relation. The so called `Dirac-like' materials have a relativistic
dispersion relation 
and typical examples include graphene [\onlinecite{grphn1, graphene2, graphene3}] and
other crystals like silicene
 [\onlinecite{sili1,sili2, gap1, gap2, gap3}], germanene [\onlinecite{STB, germanene}],  monolayer
 transition metal group-VI dichalcogenides {\mbox MX}$_2$ 
(\mbox{M=Mo, W} and \mbox{X=S, Se}) etc. [\onlinecite{gap4, gap5, TMDC, TMDC2}] which
generally have a honeycomb lattice structure. Dirac materials also have  suppressed
electron scattering and tunable electronic properties which make them very
interesting from an application point of view.

For the present study, we consider both class of systems on an equal footing
and present a unified and an exact description of 
quantum wave packet dynamics whose long term behavior displays the universal
phenomena of spontaneous collapse and 
revival. For this purpose we choose the initial localized wave-packet to be a
coherent state, which is also a minimum uncertainty wave packet, whose
cyclotron dynamics  resembles the dynamics of a classical charged particle
in a perpendicular magnetic field [\onlinecite{green}].

Our article is organized as follows:
In Sec. \ref{Sec2}  we study the wave-packet dynamics in an exact
and  unified manner for various 2D systems
with Landau levels. In addition we motivate and discuss the timescales
associated with the phenomena of 
spontaneous collapse and revival in systems with discrete and non-equidistant
energy levels. In Sec. \ref{Sec3}, we discuss the collapse 
and revival phenomenon in Schr\"odinger-like materials with parabolic
energy spectrum and $k$-linear and $k$-cubic Rashba SOI. In Sec. \ref{Sec4} we discuss
wave-packet dynamics in Dirac-like materials with a relativistic dispersion, and finally, in Sec. \ref{Sec5} we
summarize our results.

\section{Unified description of wave-packet dynamics in various 2D systems} 
\label{Sec2}
In this section we present an exact and unified formalism describing the
temporal evolution of a wave-packet in various 2D systems, in presence of
a transverse magnetic field. In particular, we focus on  both Schr\"odinger-like systems
as well as  Dirac-like materials, whose low energy properties are described by a
two band model. In Sec.~\ref{exact} we calculate the exact expectation value of the
position and velocity operator (or alternatively electric current), for an injected
coherent state minimum uncertainty wave-packet in  generic 2D systems. Next in Sec.~\ref{TS},  
we briefly review the phenomenon of wave-packet revival in various 2D systems with  Landau level spectrum.

\subsection{Exact quantum evolution of a wave packet in various 2D systems}
\label{exact}

We begin by presenting an exact unified description of the temporal evolution of the 
center of an injected wave-packet in various two dimensional systems with non-equidistant Landau energy-levels.
The Landau level eigen-spectrum describing different 2D systems can be 
written in the following generic form:

\begin{eqnarray}\label{eign_spect}
\varepsilon_n = \hbar\omega_c\Big[f(n) + \lambda\sqrt{c^2 + g(n)}\Big],
\end{eqnarray}
where $\lambda=\pm 1$ denotes two chiral energy branches, $f(n)$ and $g(n)$ are system
dependent functions of the  Landau level indexed by $n$, and $c$ is a system dependent constant term. 
Note that $f(n)$ is generally a linear function of $n$ in Schr\"odinger-like systems
arising from the parabolic part of the dispersion relation,  and it is generally
absent in Dirac-like systems.  The functions $f(n)$, $g(n)$ and $c$ for various
systems are tabulated in Table \ref{T11a}. The magnetic length, for both class
of systems,  is given by $l_c=\sqrt{\hbar/(eB)}$. The cyclotron frequency, for Schr\"odinger-like systems is given by 
$\omega_c=eB/m^\ast$ with $B$ being the applied transverse magnetic field
and $m^\ast$ is the effective mass. For Dirac-like systems $\omega_c=\sqrt{2}v_F/l_c$ where $v_F$ is the Fermi velocity.

To specify the eigen-vectors, we assume the 2D system to lie in the $x-y$ plane
and work in the Landau gauge, {\it i.e.~}the vector potential is
specified by ${\bf A} = (-By, 0, 0)$, where $B$ is the strength of
the magnetic field in the $z$-direction. The eigen-vectors corresponding
to the eigen-energies given by Eq.~(\ref{eign_spect}) with different chiralities are now given by,
\be \label{wavfnP}
\psi_{n,q_x}^+(x,y) = \frac{e^{iq_xx}}{\sqrt{2\pi a_n}}\begin{pmatrix} 
-z_n\phi_{n-m}(y-y_c)\\ \phi_n(y-y_c)\end{pmatrix}~,
\ee
and 
\be \label{wavfnM}
 \psi_{n,q_x}^-(x,y) = \frac{e^{iq_xx}}{\sqrt{2\pi a_n}}\begin{pmatrix} \phi_{n-m}(y-y_c)
 \\z_n^{*}\phi_n(y-y_c)\end{pmatrix}~,
\ee
where $m$ is an integer which depends on the related system, $|z_n|=\sqrt{g(n)}/(c+\sqrt{c^2 + g(n)})$,
$a_n=1+|z_n|^2$ and 
$\phi_n(y-y_c) = N_ne^{-(y-y_c)^2/2l_c^2}H_n((y-y_c)/l_c)$ is the harmonic oscillator 
wave function.  Other constants are given
by $N_n=\sqrt{1/(\sqrt{\pi}2^nn!l_c)}$, $y_c=q_xl_c^2$ and $H_n(x)$ denotes
the Hermite polynomial of order $n$, and 
$z_n = |z_n|$ for Dirac-like materials and $z_n = i |z_n|$ 
for Sch\"rodinger-like systems (within the chosen gauge).
Note that the eigen-system described in Eqs.~(\ref{eign_spect})-(\ref{wavfnM}) is 
applicable only for $n\ge m$. For $n<m$ there are $m$ Landau levels of `$+$' chirality with eigen-energy
$\epsilon_n^\prime=[f(n)-c]\hbar\omega_c$, and the corresponding two component eigen-vector is given by 
\be 
\psi_n({\bf r})=\frac{e^{iq_x x}}{\sqrt{2\pi}}
\begin{pmatrix} 0
 \\ \phi_n(y-y_c)\end{pmatrix}~.
\ee

\begin{table}[t] 
\begin{center} 
\caption{Landau level of various 2D systems are given by Eq.~\eqref{eign_spect}, with the following: \label{T11a}}
\begin{tabular}{ l  c c c c r} \hline \hline \\
System (Dispersion)& $f(n)$ &  $c$ &  $g(n)$ \\ \hline
2DEG with linear Rashba & $n$ & $\frac{g^\ast m^\ast}{4m_e}-\frac{1}{2}$ & $\frac{2\alpha_1^2  n}{\hbar^2\omega_c^2 l_c^2}$ \\ 
2DHG with cubic Rashba &  $n-1$& $ \frac{3g^\ast m^\ast}{4m_e}-\frac{3}{2}$ & 
$ \frac{8 \alpha_3^2 n (n-1)(n-2)}{l_c^6\hbar^2\omega_c^2} $ \\ 
Massive Dirac spectrum &$0$  &$\Delta/(\hbar \omega_c)$ & $n$  
\\ \hline
\end{tabular} 
\end{center}
\end{table}

Having described the generic form of Landau-levels and the associated eigen-vectors,
in various 2D systems with two-bands, we now turn our attention to the dynamics of an
injected wave-packet. For this purpose, we choose an initial wave-packet to be a
coherent state in a magnetic field, {\it i.e.}, a Gaussian wave packet of the following form,
\begin{equation}\label{ini_mag}
\Psi({\bf r},0)=\exp\Big({-\frac{r^2}{2l_c^2}+i q_{0}x} \Big)\frac{1}{\sqrt{\pi}l_c \sqrt{|c_1|^2 + |c_2|^2}}
\begin{pmatrix}
 c_1\\c_2
\end{pmatrix},
\end{equation}
where $\hbar q_{0}$ is the initial momentum along the $x$ direction and  the width of the 
Gaussian wave packet is considered to be equal to the magnetic length $l_c$, and the
coefficients $c_1$ and $c_2$ determine the initial spin/pseudospin polarization of the injected wave-packet. 
The idea here is to choose a minimum uncertainty wave packet, whose cyclotron dynamics
should resemble the dynamics of a classical particle in a perpendicular magnetic field. 
In addition such a wave-packet, when expressed in terms of Landau level eigen-states,
is peaked around the Landau-level $n_0 \approx l_c^2 q_0^2/2$ (see for example,
the Appendix A of Ref.~[\onlinecite{Kramer}]), and has a spread given by $\delta n = \sqrt{n_0}$. 

We note here that such a simplistic choice of the initially injected wave-packet
in Eq.~\eqref{ini_mag}, is widely used in the literature [\onlinecite{green}],  is amenable to 
analytical treatment, and gives valuable insight into the relevant timescales of the problem. 
Another realistic experimental possibility is to create wave-packets
by illuminating samples with short laser pulses [\onlinecite{laser1}].

The spinor wave packet at a later time $t$ can be written as 
\be \label{wavp_t}
\Psi_{\mu} = \int G_{\mu\nu}({\bf{r}},{\bf{r'}},t)\Psi_{\nu}({\bf{r'}},0)d{\bf{r'}}~,
\ee
where $G({\bf r},{\bf r^\prime},t)$ is the $2\times2$ Green's function matrix.
The  matrix elements of the Green's functions are defined as [\onlinecite{green}]
\be
G_{\mu\,\nu}({\bf{r}},{\bf{r^\prime}},t) = \sum_{n,\lambda} 
\int d{q_x}\psi_{n,q_x,\mu}^{\lambda}({\bf{r}},t)
{\psi_{n,q_x,\nu}^{\lambda^\ast}}({\bf{r^\prime}},0),
\ee
where $\psi_{n,q_x}^\lambda({\bf r}, 0)$  is the two component spinor
eigen-functions at $t=0$, given by Eqs. (\ref{wavfnP})-(\ref{wavfnM}).
At finite time,  $\psi_{n,q_x}^\lambda({\bf r},t)=\psi_{n,q_x}^\lambda({\bf r},0)e^{-i\epsilon_n^\lambda t/\hbar}$,
with $\epsilon_n^\lambda$ being the Landau-level energy eigen-value given in Eq. (\ref{eign_spect}).

Slightly lengthy but straightforward algebra gives the components of the Green's function matrix,  to be of the following form, 
\be \label{Grn}
G_{\mu\nu}({\bf r},{\bf r}^\prime,t) = \frac{1}{2\pi}\int_{-\infty}^{+\infty}dq_xe^{iq_x(x-x^\prime)} \eta_{\mu \nu}(q_x)~,
\ee
where $\eta_{\mu \nu}$ is given by 
\bearr
\eta_{11} &=& \sum_{n=0}^{\infty}P_{n+m}\phi_n(y-y_c)\phi_n(y^\prime-y_c)~, \\
\eta_{21} & = & \sum_{n=0}^{\infty}Q_{n+m}\phi_{n+m}(y-y_c)\phi_n(y^\prime-y_c)~, \\
\eta _{12} & = & \sum \limits_{n=0}^{\infty} R_{n+m}\phi_n(y-y_c)\phi_{n+m}(y'-y_c)~, \\
\eta_{22} & = & \sum\limits_{n=0}^{\infty} S_n\phi_n(y-y_c)\phi_{n}(y'-y_c)~,
\eearr
and we have defined $\gamma_n \equiv \omega_c \sqrt{c^2 + g(n)}$,  along with 
\bearr
P_n &=& e^{-if(n)\omega_c t}\left[e^{- i \gamma_n t} + 2 i \sin({\gamma_nt})/a_n\right]~, \\
Q_n &=& 2iz_n^{*}e^{-if(n)\omega_c t}\sin({\gamma_nt})/a_n~, \\
R_n &= & 2 i z_n e^{-i f(n) \omega_c t}\sin{(\gamma_nt)}/a_n~, \\
S_n & =& \begin{cases} e^{-i\varepsilon'_n t/\hbar}  & \mbox{for } n < m   \nonumber \\
e^{-if(n)\omega_c t}\left[e^{ i \gamma_n t} - 2 i \sin({\gamma_nt})/a_n\right] & \mbox{for } n \ge m.
\end{cases} \label{S_n} \nonumber \\
\eearr
Substituting  Eqs.~(\ref{Grn})-(\ref{S_n}) in Eq.~(\ref{wavp_t}),  we obtain the time evolved 
two component injected wave packet at a later time $t$, 
\begin{eqnarray}\label{wav_t}
\begin{pmatrix}
 \Psi_1({\bf r},t)\\ \Psi_2({\bf r},t)
\end{pmatrix}
&=&\frac{1}{\sqrt{2}\pi l_c}  \int du e^{F(x,u)}  \sum_{n=0}^\infty  (-u)^n  \frac{1}{\sqrt{c_1^2 + c_2^2}}\\
&\times& \begin{pmatrix}
\left(\frac{c_1 P_{n+m}}{2^nn!N_n} + \frac{c_2 (-u)^mR_{n+m}}{2^{n+m}(n+m)!N_{n+m}}\right) \phi_n(y-y_c)\\
\frac{c_1 Q_{n+m} \phi_{n+m}(y-y_c) + c_2 S_n \phi_{n}(y-y_c)}{2^nn!N_n} 
\end{pmatrix}, \nonumber
\end{eqnarray}
where $F(x,u)=iux/l_c-(l_c q_{0}-u)^2/2-u^2/4$ with
$u=q_x l_c$ and $\hbar q_{0}$ is the momentum of the injected wave-packet. 
We emphasize that Eq.~(\ref{wav_t}) gives the exact temporal evolution of the
coherent state injected wave function with arbitrary spin/pseudo-spin polarization, 
for a wide-class of 2D materials. 

In the rest of the paper we will focus on the specific case when the lower component
of initial wave-packet is equal to zero, {\it i.e.}, the parameters $c_1 = 1$ and $c_2 = 0$ in Eq.~\eqref{ini_mag}. 
Now, the expectation value of the position operator $\hat{\bf r}$ (or any other operator), at  time $t$, 
is simply given by 
\be \label{pos_def}
\la {\bf r}(t)\ra=\sum_{i=1}^2 \int \Psi_i^\ast({\bf r},t) \hat{\bf r}
\Psi_i({\bf r},t) d{\bf r}~.
\ee
A tedious but straightforward calculation using Eq. (\ref{wav_t}) in Eq.~(\ref{pos_def}), 
gives us the exact time dependent expectation values of the position ($x,y$), of the centre of the injected wave-packet to be 
\begin{eqnarray}\label{expX_gn}
\langle x(t)\rangle&=&l_c\sum_{n=0}^\infty\xi_n\Bigg{(}\Im\left[P_{n+m+1}(t)P_{n+m}^\ast(t)\right] \\
&+&\Im\left[ Q_{n+m+1}(t)Q_{n+m}^\ast(t) \right]\sqrt{\frac{n+m+1}{n+1}}\Bigg{)}~, \nonumber
\end{eqnarray}
and
\begin{eqnarray}\label{expY_gn}
\langle y(t)\rangle&=&l_c\sum_{n=0}^\infty\xi_n\Bigg{(}\Re[P_{n+m+1}(t)P_{n+m}^\ast(t)]
\\
&+&\Re[Q_{n+m+1}(t)Q_{n+m}^\ast(t)]
\sqrt{\frac{n+m+1}{n+1}}-1\Bigg{)}~,\nonumber
\end{eqnarray} 
where we have defined
\be \label{xi_n}
\xi_n \equiv \frac{i}{3}e^{-\frac{\tilde{q}_0^2}{3}}
\Big(\frac{-1}{12}\Big)^n\frac{1}{n!}H_{2n+1}\Big(i\sqrt{\frac{2}{3}}\tilde{q}_0\Big) ~,
\ee
with $\tilde{q} = q_0 l_c$. Note that $\xi_n$ is always real. 

The temporal evolution of an incident wave-packet which is centered around
some high energy level, in systems with discrete but non equidistant energy
levels, typically displays  the phenomena of spontaneous collapse and revival [\onlinecite{Robinett}].  However it is 
generally difficult to infer the relevant timescales associated with these
phenomena, from the exact expressions for the position operators 
given in Eqs.~(\ref{expX_gn})-(\ref{expY_gn}). Thus we briefly discuss the
phenomena of spontaneous collapse and revival, based on general arguments in the next subsection. 
Later, in Sec.~\ref{Sec3} and Sec.~\ref{Sec4},  we will show the emergence
of these timescales from the exact solution in various 2D systems.

\subsection{Oscillation, cyclotron and revival timescales}
\label{TS}
Discretized Landau energy-levels are formed whenever a 2D electronic system,
is subjected to a strong  perpendicular magnetic field. The spatio-temporal evolution of wave-packets in  such quantum 
system with discrete but non-equidistant 
energy spectrum is generally quiet complex and exhibits both classical and
quantum behavior. The quantum behavior being manifested in the form of
spontaneous collapse and long term quantum revival of the wave-packet
arising due to quantum interference.  However  well defined periodicities for 
quasi-classical behavior, spontaneous collapse and revival
emerge [\onlinecite{Nauenberg, Robinett, Romera_PRB, Demikhovskii_PRA}],
if the initial wave packet have a substantial overlap with some large Landau level denoted by $n_0$.

Various timescales during the wave packet dynamics in a discrete system with
non-equidistant energy spectrum,  can be inferred from the analytic form of
the autocorrelation function [\onlinecite{Nauenberg}] of the wave packet, which is defined 
as $A(t) = \langle \Phi({\bf{r}},t) |\Phi({\bf{r}},0) \rangle$. Expanding $\Phi({\bm r}, t)$ in terms of the orthonormal
eigenstates of the system under consideration, $\{\phi_n\}$,  we get
$ \Phi({\bf{r}},t) = \sum_n c_n \phi_n({\bf r}) e^{- i \epsilon_n  t/\hbar}$, where $\epsilon_n$ are the discrete
energy eigenvalues of the system and $c_n = \langle \phi_n ({\bf{r}})| \Phi({\bf{r}}, 0)\rangle$. The autocorrelation 
function is now given by  
 \be
 A(t) =  \sum_n |c_n|^2 e^{i \epsilon_n  t/\hbar}~.
\ee 
For studying the dynamics of a  localized injected wave packets,  which can be
expressed as a superposition of the eigen-states of the system  centered around
some large Landau-level  $n=n_0$, such that $n_0 \gg \delta n  \gg 1$,
we can assume the form of the expansion coefficients to a Gaussian centered around $n_0$ with spread of $\delta n$. 
Doing a Taylor series expansion of the
energy, $\epsilon_n  = \epsilon_{n_0} + (n-n_0) \epsilon_{n_0}' + (n-n_0)^2 \epsilon_{n_0}''/2 + \dots $,
where $\epsilon_n' = (d\epsilon_n/dn)_{n=n_0} $ and so forth, the autocorrelation function  can be rewritten as, 
\be \label{eq:auto}
A(t) = \sum_{n = - \infty}^{\infty} |c_n|^2 e^{i t/\hbar \left( \epsilon_{n_0} + (n-n_0) \epsilon_{n_0}' 
+ (n-n_0)^2 \epsilon_{n_0}''/2 + \dots \right) }~.
\ee
The coefficients of the Taylor expansion of the energy $\epsilon_n$ in the
exponential of Eq.~\eqref{eq:auto} defines a characteristic timescale via, 
\be \label{timescale1}
\tau_{\rm osc} = \frac{2 \pi \hbar}{\epsilon_{n_0}} ~,~~~ \tau_{\rm cl} = \frac{2 \pi \hbar}{|\epsilon_{n_0}'|}~,
~{\rm and}~~~ \tau_{\rm rev} = \frac{4 \pi \hbar}{|\epsilon_{n_0}''|}~.
\ee
The timescale $\tau_{\rm osc}$, is an intrinsic  quantum oscillation time scale,
which does not lead to any quantum interference in the various wave packet components
of different Landau-levels and is thus it does not effect the long term 
dynamics of the system. At the { `classical' cyclotron} time-scale $\tau_{\rm cl}$,
the wave-packet evolves quasi-classically and the center of the wave-packet completes
one cyclotron orbit, and returns to the initial
position and the autocorrelation function approximately reaches its initial value. 

At larger timescales quantum interference between the wave function components of
different Landau levels  in the incident wave-packet, leads to spontaneous collapse
of the wave function and then to quantum revival. 
The quantum revival of the wave packet, over timescale $ \tau_{\rm rev}$, occurs due
to constructive interference when the terms proportional to the second derivative in
the energy are in multiples of $2 \pi$. This revival hierarchy sustains even in the  higher order terms.
In addition the time at which the spreading of the wave packet leads to quantum
self-interference which leads to spontaneous collapse, is
given by $\tau_{\rm coll} = \tau_{\rm rev}/({\delta n})^2$. See Ref.~[\onlinecite{Robinett}] for a detailed review. 

For this article, where the generic Landau-level spectrum for various 2D systems
is given by Eq.~\eqref{eign_spect}, corresponding derivatives of the energy
with respect to $n$, which appear in Eq.~(\ref{timescale1}) are explicitly given by 
\be
\frac{\varepsilon_n'}{\hbar\omega_c} = f'(n) +\lambda \frac{g'(n)}{2 \sqrt{c^2 + g(n)}}~,
\ee
and 
\be
\frac{\varepsilon_n''}{\hbar\omega_c} = f''(n)+ 
\frac{\lambda}{2 \sqrt{c^2 + g(n)}}\Big{[}g''(n)-\frac{g'(n)^2}{2(c^2 + g(n))}\Big{]}, 
\ee
where $f(n)$ is typically a linear function of $n$,  arising from the parabolic
part of the dispersion relation, which exists only for Schr\"odinger-like systems,
and thus $f'(n)$ is a constant and $f''(n) = 0$.   

The classical, spontaneous collapse and revival timescales for various 2D systems
studied in this paper, along with the relevant material parameters,
is tabulated in Table~\ref{T2} and Table~\ref{T3}. We now proceed to
study the phenomena of revival and collapse in various systems like $k$-linear Rashba 2DEG, $k$-cubic rashba 2DHG etc. 

\section{Schr\"odinger-like systems}
\label{Sec3}
In this section we  discuss spontaneous 
decay  and long-term quantum revival of a quantum wave packet 
in two-dimensional Schr\"odinger-like fermionic systems 
described by Rashba spin-orbit interaction (SOI).
Let us first consider the case  of a 2DEG with  $k$-linear Rashba SOI.

\subsection {2DEG with $k$-linear Rashba SOI} 

\begin{figure}[t]
\begin{center} 
\includegraphics[width=1.0 \linewidth]{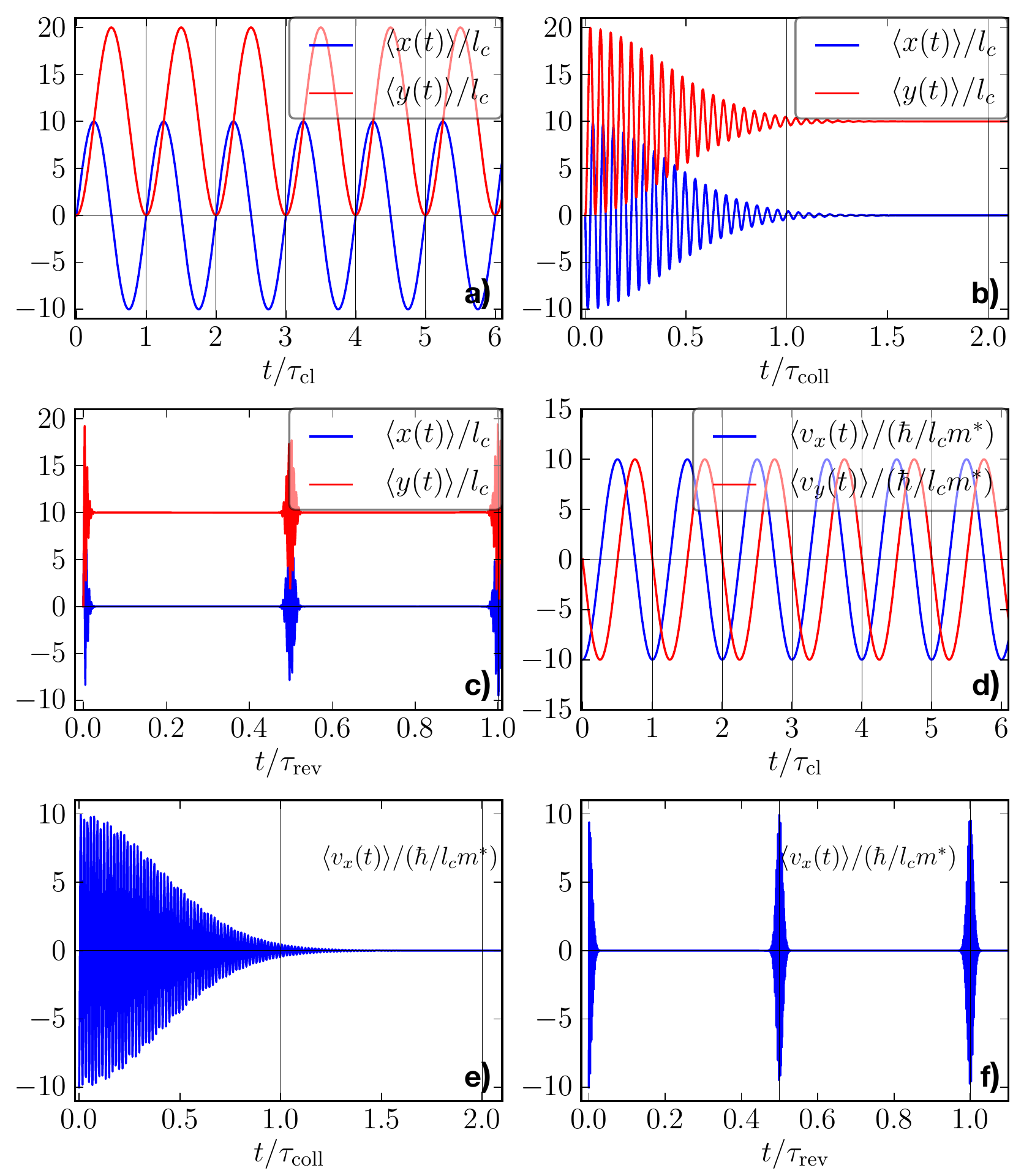}
\end{center}
\caption{Panel a), b), and c) show the position of the center of the wave
packet vs time, while panels d), e) and f) display the velocity expectation
value,  over different timescales, {\it i.e.~}$\tau_{\rm cl}$, $\tau_{\rm coll}$ and $\tau_{\rm rev}$ respectively. 
Note that in panels a), b) and c)  $\langle x(t)\rangle$ 
is centered around 0, whereas $\langle y(t)\rangle$ is centered around $y_c=q_0l_c^2 = 10$. 
Here we have chosen $B=2$T and other material parameters for AlGaAs/GaAs quantum well: 
$m^\ast=0.067m_e$, $g^\ast=-0.44$, $\alpha_1=10^{-13}$eV-m.}
\label{fig1}
\end{figure}

We now examine the wave-packet dynamics in a 2DEG formed at the interface of
an {\it inversion asymmetric} III-V semiconductor quantum well, subjected to a Zeeman field 
perpendicular to the interface. The inversion asymmetry of the quantum well
gives rise to Rashba SOI [\onlinecite{rashba}], which has a linear dependence on momentum.

To obtain the Landau level spectrum of a Rashba 2DEG in a transverse magnetic field, we work with 
Landau gauge ${\bf A}=(-By,0,0)$ for the vector potential. Making the Landau-Peierls substitution  
${\bf p} \to {\bf \Pi} = {\bf p}+e{\bf A}/\hbar$, the Hamiltonian [\onlinecite{ras_mg}] describing Rashba 2DEG is given by 
\be \label{Ham_lin}
H=\frac{{\bf \Pi}^2}{2m^\ast}+\frac{\alpha_1}{\hbar}
\Big(\sigma_{x}\Pi_{y}-\sigma_{y}\Pi_{x}\Big)
+\frac{1}{2}g^\ast \mu_B \sigma_z B~,
\ee
where $m^\ast$ is the effective mass of the electron,
$\alpha_1$ is the Rashba SOI coupling coefficient, $g^\ast$ is the effective
Lande g-factor, $\mu_B$ is the Bohr magneton and $\sigma_i$ denote the
Pauli spin matrices.
The Landau-level  eigen-spectrum corresponding to Eq.~(\ref{Ham_lin}) is  
given [\onlinecite{ras_mg}] by the generic form considered in Eq.~(\ref{eign_spect})
with the following substitutions: 
\be
f(n) = n~,~~ c = \frac{g^\ast m^\ast}{4m_e}-\frac{1}{2}~, ~~{\rm and}~~ g(n) = \frac{2\alpha_1^2  n}{\hbar^2\omega_c^2 l_c^2}~,
\ee
where $m_e$ is the free electron mass.
The corresponding eigen-vectors 
for the two spin-split branches ($\lambda = \pm 1$) are now simply obtained by putting 
$m=1$ in Eqs.~(\ref{wavfnP})-(\ref{wavfnM}).

Now the temporal evolution of the expectation value of the coordinates $(x,y)$ of
the centre of the injected Gaussian wave-packet are given by
Eq.~(\ref{expX_gn}) and Eq.~(\ref{expY_gn}) with $m \to 1$. As a check of
our calculation, we note that the expectation values of the position operator
are consistent with that derived in Eqs.~(36a)-(36b) of Ref.~[\onlinecite{green}].

Let us now calculate the time-dependent expectation values of the velocity operator,  which also gives the charge current. 
Using the Heisenberg equation of motion for the position operator, 
$\hat{\bf v}= -i \hbar^{-1} [\hat{\bf r}, H]$, the components of the 
velocity operator are given by 
\be
\hat{v}_x=\frac{\hat{\Pi}_x}{m^\ast}-\frac{\alpha_1}{\hbar}\sigma_y~, 
~~\quad ~{\rm and~} \quad \hat{v}_y=\frac{\hat{\Pi}_y}{m^\ast}+\frac{\alpha_1}{\hbar}\sigma_x~.
\ee
Following the same procedure as described in Sec. \ref{exact},
we finally obtain the following expressions for the
components of the velocity expectation values,
\be
\begin{pmatrix} \la v_x(t)\ra \\ \la v_y(t)\ra \end{pmatrix}
= \frac{\hbar}{l_c m^\ast}
\sum_{n=0}^\infty \xi_n  \begin{pmatrix} \Re v_n \\ \Im v_n \end{pmatrix}~,
\ee
where we have defined
\be
v_n \equiv P_{n+2}^\ast P_{n+1} + Q_{n+2}^\ast Q_{n+1} \sqrt{\frac{n+2}{n+1}} 
+ \frac{i\tilde{\alpha}_1}{\sqrt{n+1}} P_{n+2}^\ast Q_{n+1}~,
\ee
using the dimensionless SOI strength: $\tilde{\alpha}_1 = \sqrt{2} \alpha_1/(l_c \hbar \omega_c)$
and $\xi_n$ is defined in Eq.~(\ref{xi_n}). 

To discuss wave-packet dynamics which shows the phenomena of spontaneous collapse  and revival, 
we compute the relevant time scales described in Sec.~\ref{TS} for 2DEG with $k$-linear Rashba SOI.
The oscillation, classical and quantum revival timescales for this system are respectively given by
\be
\tau_{\rm osc}^\lambda=\tau_c\frac{1}{\lvert n_0+\lambda \sqrt{c^2+\tilde{\alpha}_1^2 n_0} \rvert}~,
\ee
\begin{eqnarray}
\tau_{\rm cl}^\lambda=\tau_c\frac{1}{\lvert 1+\lambda\frac{\tilde{\alpha}_1^2}{2\sqrt{c^2+\tilde{\alpha}_1^2 n_0}}\rvert},
\end{eqnarray}
and finally, 
\begin{eqnarray}
\tau_{\rm rev}=\tau_c
\frac{8[{c^2+\tilde{\alpha}_1^2 n_0}]^{3/2}}{\tilde{\alpha}_1^4}, 
\end{eqnarray}
where $\tau_c=2\pi/\omega_c$.
Note that there are two contributions in $\tau_{\rm osc}$ and $\tau_{\rm cl}$ coming from the 
upper and lower branches of the energy spectrum. In the rest of the paper, we will be using the average timescales
defined as $\tau_{\rm osc}=(\tau_{\rm osc}^++\tau_{\rm osc}^-)/2$ and $\tau_{\rm cl}=(\tau_{\rm cl}^++\tau_{\rm cl}^-)/2$.

Experimentally, $k$-linear Rashba SOC is present in an AlGaAs/GaAs
quantum well [\onlinecite{Gossard}] or in an InGaAs/InAlAs quantum well [\onlinecite{Nitta}]
among other materials. 
Various parameters and the corresponding timescales for these are given in Table.~\ref{T2}. 
We study the dynamics of the center of the wave-packet and its velocity
(or equivalently current), in Fig.~\ref{fig1}, using the material parameters of an AlGaAs/GaAs
quantum well.  Both the position and velocity of an initial minimum-uncertainty wave-packet display
the phenomena of spontaneous collapse and consequent full quantum revival at long times.

\subsection {2D systems with $k$-cubic Rashba SOI}

\begin{figure}[t]
\begin{center} 
\includegraphics[width=1.0 \linewidth]{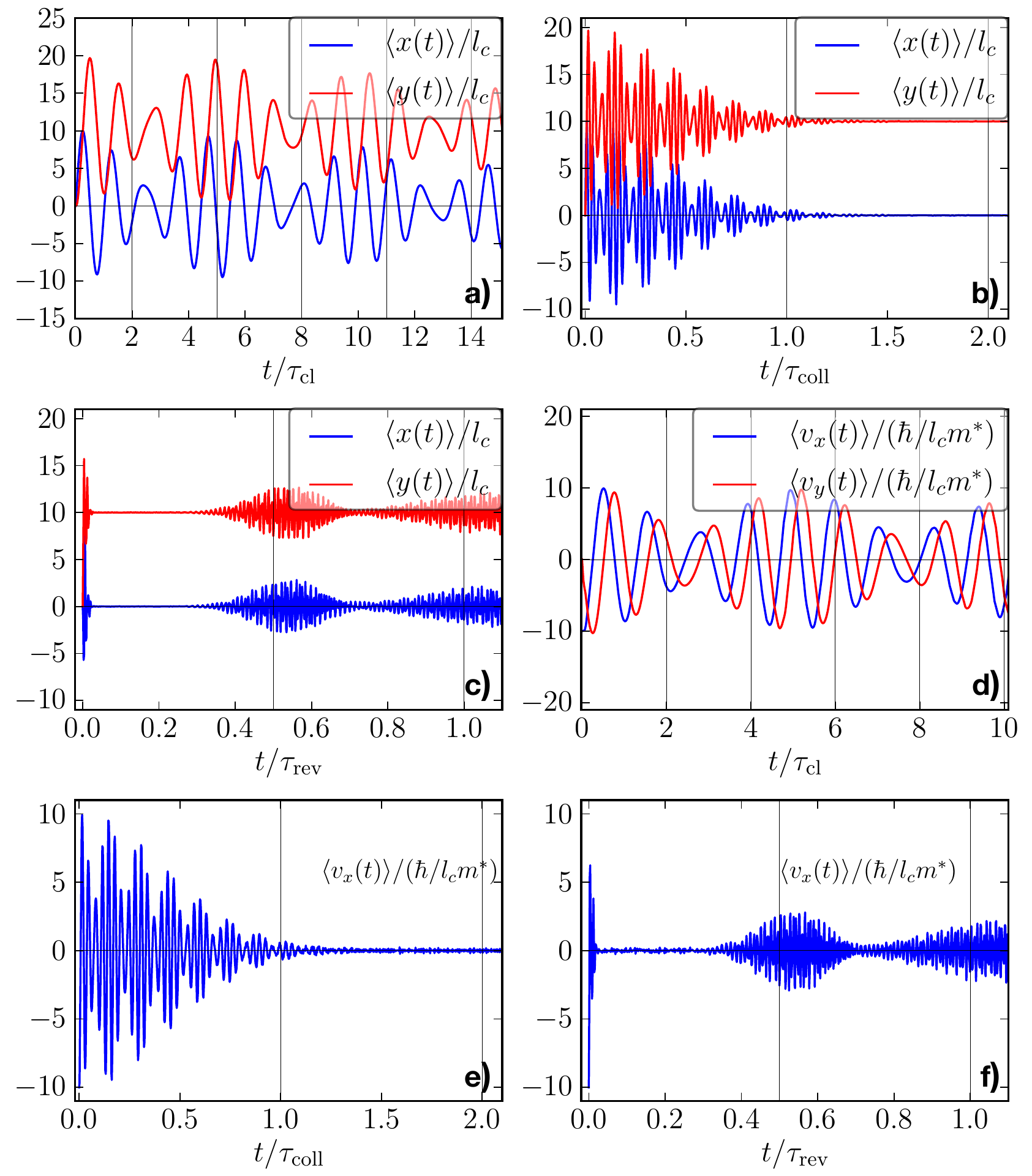}
\end{center}
\caption{As in Fig.~\ref{fig1}, panels a), b), and c) show the position, and panels d),e), and f)
show the  velocity expectation vs time, for different timescales. 
Here we have chosen $B=2$T and other material parameters for 2D hole gas formed at $p$-type 
GaAs/AlGaAs heterostructure which are:  
$m^\ast=0.45m_e$, $g^\ast=7.2$, $\alpha_3=10^{-29}$eV-m$^3$.}
\label{fig2}
\end{figure}

\begin{table*}[t] 
\begin{center} 
\caption{The parameters and associated timescales for various 
2D Schr\"odinger-like systems. Here $n_0=50$ and $B=2$T.
 \label{T2}}
\begin{tabular}{ l c c c c c c  r} \hline \hline
System & material & $m^\ast/m_e$ &  $g^\ast$ & SOI:  $\alpha_1$(eV-m) & $\tau_{\rm osc}$ & $\tau_{\rm cl}$ & $\tau_{\rm rev}$\\ 
     \   &(quantum well)&         &           & or \  $\alpha_3$ (eV-m$^3$)  & (fs)& (ps) & (ns) \\ \hline
\vspace{0.1cm}      
Systems with \\ linear Rashba  SOI & AlGaAs/GaAs \cite{Gossard}  & 0.067 & -0.44 &
$\alpha_1 =  10^{-13}$& 23.94 & 1.19 & 48.3$\times10^6$\\ 
	& InGaAs/InAlAs \cite{Nitta} & 0.052 & 4.0 & $\alpha_1 =10^{-11}$  &18.59 & 0.93 & 18.04 \\

\vspace{0.1cm}
2DEG with \\ cubic Rashba SOI & GaAs/AlGaAs (2DHG)\cite{hole1, Winkler, hole5} & 0.45 & 7.2&
$\alpha_3 = 10^{-29}$& 165 & 8.11 &  13.70\\ 
			 & SrTiO$_3$ (2DEG)\cite{Sr1,Sr3} & 1.45 & 2.0 & $\alpha_3 = 10^{-30}$  & 529 & 25.92 & 102.02  \\
\\ \hline
\end{tabular} 
\end{center}
\end{table*}

In this section we investigate the wave-packet dynamics in  2DEG and 2DHG with parabolic
dispersion relation along with the Rashba SOI which is cubic in momentum. Generally $k$-cubic 
Rashba SOI occurs in two different systems. One of such systems is 
2D heavy hole gas [\onlinecite{hole1,Winkler,hole3,hole4, hole5}] formed at the interface of $p$-doped III-V semiconductors, 
namely {\mbox GaAs}/{\mbox AlGaAs} heterostructure. In addition $k$-cubic Rashba SOI can also be 
found in the 2D electron gas formed at the interface of perovskite oxide structures [\onlinecite{Sr1,Sr2, Sr3}] such as
{\mbox LaAlO}${}_3$/{\mbox SrTiO}${}_3$ interface and {\mbox SrTiO}${}_3$
surface.

The single particle Hamiltonian [\onlinecite{tianx,zarea}] 
of a 2D system  with cubic Rashba SOI, in a transverse magnetic field
is given by 
\begin{eqnarray}\label{Ham_cub}
H=\frac{{\bf \Pi}^2}{2m^\ast}+\frac{i \alpha_3}{2\hbar^3}
\Big(\Pi_{-}^3\sigma_{+}-\Pi_{+}^3\sigma_{-}\Big)
+ \frac{3}{2}g_s \mu_B {\bs \sigma} \cdot {\bf B},
\end{eqnarray}
where ${\bf \Pi}$ is the conjugate momentum defined in previous section,
$\alpha_3$ is the Rashba coupling coefficient.  Additionally, we have defined
$\Pi_{\pm} \equiv \Pi_x\pm i\Pi_y$, and $\sigma_\pm \equiv \sigma_x\pm i\sigma_y$. 
Note that for the case of heavy holes,  the Pauli matrices represent an effective pseudo-spin with 
spin projection $\pm3/2$ along the growth direction of the quantum well.

The Landau level spectrum is again given by Eq.~\eqref{eign_spect}, with 
\be
f(n)=n-1~, ~~~ g(n) =  \frac{8 \alpha_3^2 n (n-1)(n-2)}{l_c^6\hbar^2\omega_c^2}~, 
\ee 
and finally 
\be
c= \frac{3g^\ast m^\ast}{4m_e}-\frac{3}{2}~.
\ee
Similarly, the temporal evolution of the expectation values of the position operators
for the centre of the wave-packet are given by substituting
$m=3$ in Eqs.~(\ref{expX_gn})-(\ref{expY_gn}).
The components of the velocity operator are given by
\be
\hat{v}_x=\frac{\Pi_x}{m^\ast}\sigma_0+\frac{3i\alpha_3}{2\hbar^3}
(\sigma_{+}\Pi_{-}^2-\sigma_{-}\Pi_{+}^2)~,
\ee
and
\be
\hat{v}_y=\frac{\Pi_y}{m^\ast}\sigma_0+\frac{3\alpha_3}{2\hbar^3}
(\sigma_{+}\Pi_{-}^2+\sigma_{-}\Pi_{+}^2)~.
\ee
Their expectation values are given by 
\be
\begin{pmatrix} \la v_x(t)\ra \\ \la v_y(t)\ra \end{pmatrix}
= \frac{\hbar}{l_c m^\ast}
\sum_{n=0}^\infty \xi_n  \begin{pmatrix} \Re h_n \\ \Im h_n \end{pmatrix}~,
\ee
where we have defined
\begin{eqnarray}
h_n &\equiv& P_{n+4}^\ast P_{n+3}  +  \sqrt{\frac{n+4}{n+1}}  Q_{n+4}^\ast Q_{n+3} \nonumber\\
&+ &3 i \tilde{\alpha}_3 \sqrt{\frac{(n+2)(n+3)}{n+1}} P_{n+4}^\ast Q_{n+3},
\end{eqnarray}
using the dimensionless SOI strength: 
$\tilde{\alpha}_3 = 2\sqrt{2} \alpha_3/(l_c^3 \hbar \omega_c)$ and $\xi_n$ is defined in Eq.~(\ref{xi_n}). 
 Note that the expectation values of the position and velocity operators for systems
 with $k$-cubic Rashba SOI have recently been reported in Ref.~[\onlinecite{tutul}],
 and are presented here  for completeness, and to emphasize that they can be derived from our unified description.

The oscillation, classical, and revival timescales are now given by, 
\be
\tau_{\rm osc}^\lambda=\tau_c\frac{1}{\lvert n_0-1+\lambda \sqrt{c^2+g(n_0)}\rvert}~,
\ee
\be
\tau_{\rm cl}^\lambda=\tau_c\frac{1}{\lvert 1+\lambda\frac{\tilde{\alpha}_3^2(3n_0^2-6n_0+2)}{2\sqrt{c^2+g(n_0)}}\rvert}~,
\ee
and
\be
\tau_{\rm rev}=\tau_c
\frac{4\sqrt{c^2+g(n_0)}}{\tilde{\alpha}_3^2\Big\{6(n_0-1)-\frac{\tilde{\alpha}_3^2}{2[c^2+g(n_0)]}(3n_0^2-6n_0+2)^2\Big\}}~. \\ 
\ee

The phenomena of spontaneous collapse is evident in Fig.~\ref{fig2}b and Fig.~\ref{fig2}e for
the position and the velocity of the centre of the 
injected wave-packet respectively. The phenomena of partial and full quantum revival in
2DHG is evident in Fig.~\ref{fig2}c and Fig.~\ref{fig2}f.

\section{Dirac materials}
\begin{figure}[t!]
\begin{center} 
\includegraphics[width=1.0 \linewidth]{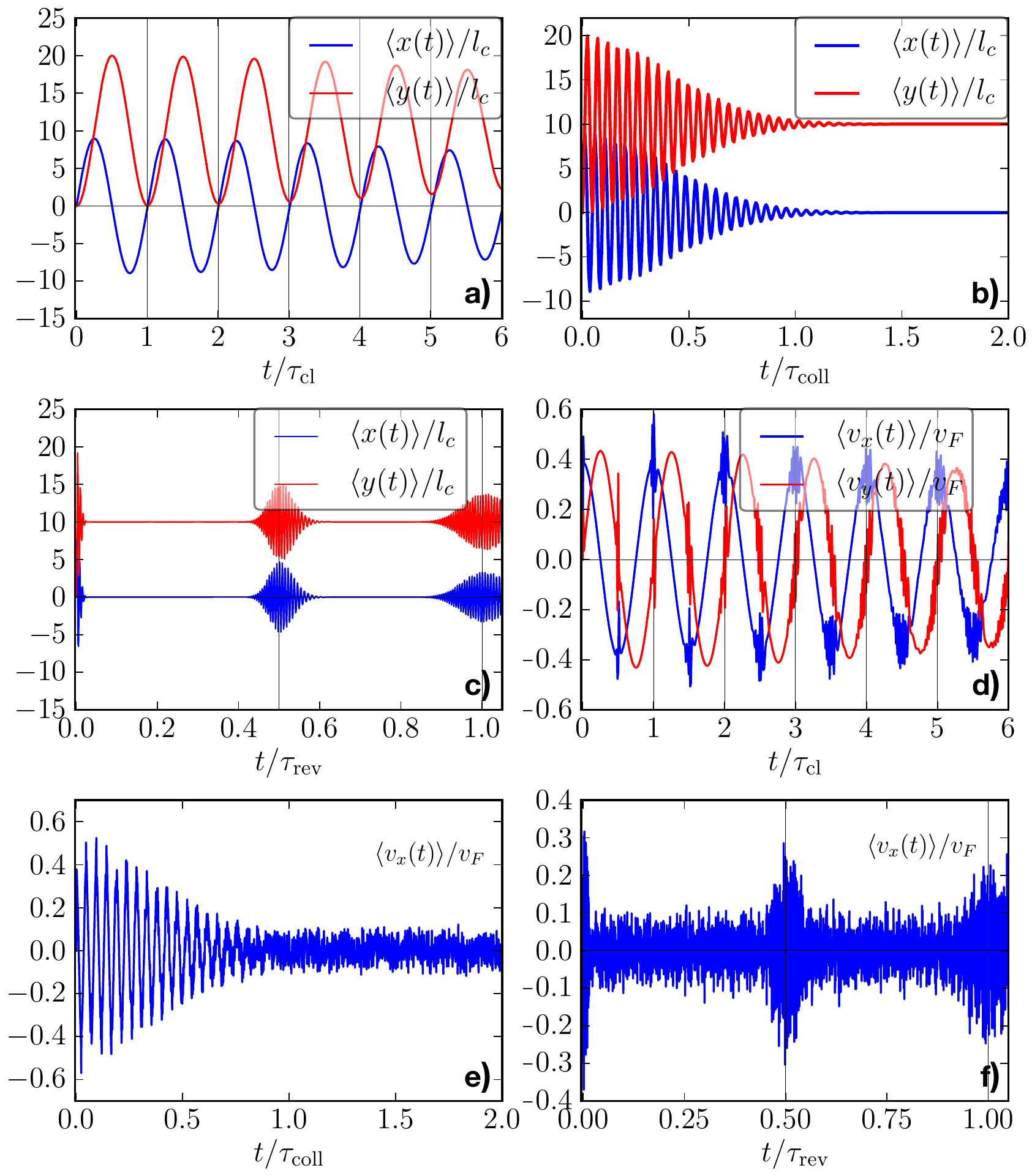}
\end{center}
\caption{As in Fig.~\ref{fig1}, panels a), b), and c) show the position, and panels d),e), and f) show
the  velocity expectation vs time, for different timescales. 
Other parameters are $\Delta=0.4$eV, $v_F=532000$ m/s, $B = 2$T
 and $q_0l_c$=10.}
\label{fig3}
\end{figure}

\label{Sec4}
\begin{table*}[t]
\caption{The parameters for various 2D Dirac-like materials and the associated time-scales. Here  $n_0=50$ and $B=2$T.
\label{T3}}
\vspace{0.3cm}
\begin{tabular}{l c c c c c  r } \hline\hline
Material & Fermi velocity & Band Gap & $\tau_{\rm osc}$ & $\tau_{\rm cl}$& $\tau_{\rm rev}$ \\
         & ( $10^6$ ms$^{-1})$ & (eV) &   (fs) & (ps) & (ns) \\ \hline
\vspace{0.1cm}
Graphene\cite{grphn1, graphene2, graphene3}  & 1 & Gapless &  11.39 & 1.13 & 0.22 \\ 
\vspace{0.1cm}
Silicene\cite{gap1,gap2,gap3} & 0.532 & 0 - 0.2 (tunable) &  21.4-19.02 & 2.14-2.41 & 0.42-0.61\\
\vspace{0.1cm}
Germanene \cite{STB} & 0.517 & 0 - 0.2 (tunable) & 22.0-19.4 & 2.2-2.5 & 0.44-0.64\\
\vspace{0.1cm}
MoS$_2$ \cite{gap4,gap5}& 0.085 & 1.6 &  5.16 & 348.12 & 4.69$\times10^4$\\ \hline
\end{tabular}
\end{table*}

In this section we focus on systems whose energy dispersion is similar to that given by the
relativistic Dirac equation,
such as graphene [\onlinecite{grphn1, graphene2, graphene3}], silicene  [\onlinecite{sili1,sili2, gap1, gap2, gap3}],
monolayer group-VI dichalcogenides {\mbox MX}$_2$ 
(\mbox{M=Mo, W} and \mbox{X=S, Se}) etc [\onlinecite{gap4, gap5, TMDC, TMDC2}].
The low-energy Hamiltonian describing these systems is simply given by, 
\be \label{Hdirac}
 H = v_F(p_x\sigma_x+p_y\sigma_y) + \Delta\sigma_z,
\ee
where $v_F$ is the Fermi velocity, and $2 \Delta$ is the band 
gap, both of which differ from system to system (see Table~\ref{T3}). If the
band gap $\Delta \to 0$, then Eq.~\eqref{Hdirac} describes 
massless Dirac fermions as in graphene. Typically,  the band gap varies
from $0-1$ eV in these systems, and in silicene it can even be tuned
experimentally by means of an externally applied electric field [\onlinecite{STB}]. 

The effect of a transverse magnetic field is included by the usual Landau-Peierls substitution  
 in Eq.~\eqref{Hdirac}.
The Landau-Level eigen-spectrum is again given by the generic Eq.~\eqref{eign_spect}, after substituting 
\be 
f(n)=0~,~~~  g(n)= n,~~~ {\rm and}~~~ c= \Delta/(\hbar \omega_c)~. 
\ee  
The corresponding Landau level wave-functions are  given by Eqs.~\eqref{wavfnP}-\eqref{wavfnM}, with the substitution $m \to 1$. 

The time-evolved injected wave-packet at a later time $t$ can be obtained from Eq.~\eqref{wav_t}, with $m=1$.
The coordinates of the center of the wave packet at later times are now given
by the generic Eqs.~\eqref{expX_gn}-\eqref{expY_gn} after the substitution $m \to 1$. 
The velocity operator is given by the Heisenberg equation of motion: $\hbar \hat{v}_j = i [\hat{H}, {\hat r}_j]$, and
straightforward calculations yield the following expressions for the expectation value
for the velocity of the center of the injected wave-packet,
\be \label{velB1}
\langle v_x(t)\rangle = \Re [\la v(t) \ra]~,~~{\rm and},~~\la v_y(t)\ra = \Im [\la v (t)\ra]~,
\ee
where 
\be \label{velB2}
\la v(t) \ra = \sqrt{2} v_F \sum_{n=0}^\infty \xi_n \frac{1}{\sqrt{1+n}} ~P_{n+2}^\ast Q_{n+1}~.
\ee

The oscillation, classical and revival timescales in this case are given by,
\be\label{timescale2}
\tau_{\rm osc} = \frac{2\pi\hbar}{\varepsilon_{n_0}}~,~~~ \tau_{\rm cl} = \frac{4\pi\hbar\varepsilon_{n_0}}{\epsilon^2}~,
 ~~{\rm and} ~~\tau_{\rm rev} = \frac{16\pi\hbar\varepsilon_{n_0}^3}{\epsilon^4}~,
\ee
where $\varepsilon_{n_0}=\sqrt{\Delta^2+\epsilon^2n_0}$ and $\epsilon=\sqrt{2}\hbar v_F/l_c.\vspace{0.5cm}$

We tabulate different materials, whose low energy properties are described by Eq.~\eqref{Hdirac} in
Table~\ref{T3}, along with their material properties, and the 
relevant oscillation, classical and revival timescales. We plot the expectation value
of the position and velocity of the center of the wave-packet in Fig.~\ref{fig3} over the
classical, collapse and revival timescales to highlight the the phenomena of spontaneous collapse and quantum revival.  

Note  that the value for band gap plays an important role in determining the relevant timescales. In particular the revival time
for MoS$_2$ is quite large as compared to other Dirac-like materials due to its large band-gap. However the band-gap is not a fixed 
quantity, and in some materials it can be varied by the application of a transverse electric fields (as in silicene) or by doping 
(as in transition Metal dichalcogenides).

\section{Summary and conclusions}
\label{Sec5}
We present a unified treatment of wave-packet dynamics in various 2D systems,
in presence of a transverse magnetic field, and the associated phenomena of 
spontaneous collapse and long term quantum revival of the wave-packet.
In particular we focus on a  minimum uncertainty Gaussian wave packet
and obtain exact expressions for the expectation values of the position
and velocity operators for a variety of 2D materials, in addition to the
various timescales associated with the  phenomena of spontaneous collapse and quantum revival. 

For any system with discrete and non-equidistant Landau level spectrum
injecting an initial electron wave packet which is peaked (centered) around some Landau level, we
find that the wave packet initially evolves quasi-classically and oscillate with a period of $\tau_{\rm cl}$ 
(the cyclotron time-period). However at larger times, the wave packet
eventually spreads and quantum interference between the different Landau
level components of the wave-function leads to its `collapse', and at even longer times, 
that are multiples (or rational fractions) of $\tau_{\rm rev}$, quantum 
interference results in long-term revival of the wave packet, and the electron 
position and velocity regains its initial amplitude --- again undergoing quasiclassical oscillatory motion.

To summarize finally, we present a unified and exact analytical treatment of
wave-packet dynamics in various 2D spin-orbit coupled Schr\"odinger systems
and Dirac-like materials. As expected on general grounds, for any system with
phenomenon of spontaneous collapse and quantum revival.

\section*{Acknowledgements}  
A. A. gratefully acknowledges funding from the INSPIRE Faculty Award by DST  
(Govt. of India), and from the Faculty Initiation Grant by IIT Kanpur, India.

\end{document}